\documentclass[
    ,final            
  ]
  {aipproc}

\layoutstyle{6x9}

\begin{document}

\title{Sunyaev-Zeldovich Observations of Massive Clusters of Galaxies}

\author{P. Gomez}{
  address={Gemini Observatory, 670 N A'Ohoku Place, Hilo, HI 96720, USA},
email={pgomez@gemini.edu}
}

\author{A. K. Romer, J. B. Peterson, W. Chase}{
  address={Carnegie Mellon University, 5000 Forbes Avenue, Pittsburgh, PA 15213, USA}
}

\author{M. Runyan}{
  address={University of Chicago, 5640 South Ellis Avenue, Chicago, IL 60637, USA}
}

\author{W. Holzapfel, C. L. Kuo, M. Newcomb}{
  address={University of California at Berkeley, 366 LeConte Hall, 730, Berkeley, CA 94720, USA}
}

\author{J. Ruhl, J. Goldstein}{
  address={Case Western Reserve University, 10900 Euclid Avenue, Cleveland, OH 44106, USA}
}

\author{A. Lange}{
  address={Caltech, Mail Code 509-33, Pasadena, CA 91125, USA}
}

\begin{abstract}
 	We present detections of the Sunyaev-Zeldovich Effect (SZE) at 150GHz and 275GHz for the X-ray luminous z=0.299 cluster 1E0657-67. These observations were obtained as part of an X-ray, weak lensing, and SZE survey of nearby X-ray bright clusters. The SZE 
maps were made with the ACBAR (150, 210, 275 GHz) bolometer array installed at the Viper telescope located at the South Pole. We also present preliminary results from a blind SZE cluster survey.
\end{abstract}

\maketitle


\section{Introduction}

 We are undertaking a multiwavelength study of the physical properties of a sample of X-ray bright galaxy clusters. These clusters were selected from the REFLEX survey of southern X-ray clusters (~\cite{2001A&A...369..826B}). They conform to the
following criteria: L$_x > 4\times 10^{44}$ erg/s [0.1-2.5 keV] and $\delta < -44^{\circ}$. In this study, we will combine new and
archival X-ray data with weak lensing observations and new Sunyaev-Zeldovich Effect (SZE) measurements.

The rationale behind combining these multiwavelength observation is the fact that they
probe the cluster components in different ways. For instance, X-ray and SZE data probe the gas density and temperature distributions, whereas 
weak lensing observations probe the total mass distribution. The 
combination of
X-ray and SZE data provides three observables (X-ray surface
brightness, SZE decrement, and projected temperature) which constrain
the variation of density and temperature throughout the target
cluster. Comparing weak-lensing mass distributions with X-ray and
SZE gas distributions in the same clusters allows a detailed test
of usual assumptions (hydrostatic equilibrium, spherical symmetry, and
the lack of significant substructure) on which estimates for the
amount and distribution of dark matter in clusters are based.  

This study will attempt to:  1) characterize 
the density, temperature, and dynamical state
of the gas in each cluster. 2) compare the mass estimates derived from the SZE and X-ray 
observations with the more direct 
weak lensing mass. 
This study 
is paramount to determine whether, and to what level, future cosmological
parameter estimates from SZE and X-ray cluster surveys might be biased by non-thermal
cluster physics and/or by non standard cluster geometries (e.g., \cite{2001ApJ...560L.111H}).

\subsection{Sunyaev-Zeldovich Effect Observations with ACBAR}

	The SZE describes the inverse Compton scattering of Cosmic Microwave Background (CMB) radiation
photons by hot electrons in the intra-cluster plasma.  There are two
SZ effects: the kinetic and thermal effects.  The thermal effect is due
to the random thermal motion of the cluster electrons and shifts lower
energy CMB photons to higher energy while the kinetic effect is due to
the bulk flow motion of the cluster. Since photon number is conserved
in the process, the net effect of the thermal effect is to distort the
CMB spectrum such that there is a null at about 215 GHz, an increment at shorter wavelengths, and a decrement at longer wavelengths. This spectral distortion allows
multi-frequency instruments that span this null to separate the SZE
from intrinsic CMB anisotropies. Except at frequencies close to the thermal null, the thermal effect dominates the kinetic effect by an order of magnitude.

	We have performed cluster SZE observations with the Arcminute Cosmology Bolometer Array Receiver (ACBAR) mounted at the Viper telescope (~\cite{2000ApJ...532L..83P}). This millimeter/submillimeter telescope consists of 4 mirrors
arranged in an off-axis configuration. The radiation from an object collected by the 2.15m diameter primary and the secondary (which form an aplanatic Gregorian system) is reflected by a flat chopping tertiary mirror and focused by a hyperbolic condenser onto the focal plane of the telescope. Note that the tertiary mirror is located at the exit pupil of the system. Therefore, any motion of the chopping mirror is equivalent to a motion of the primary mirror. This allows us to perform fast raster scans of up to 3 $^{\circ}$ in azimuth at a fixed elevation.

The ACBAR instrument (\cite{run1}) consists of 16 micro-lithographed
`spider-web' bolometers arranged in a $4\times4$ array.  The detectors are
spaced by $\sim$ 16 arcmin and have beams that can be modeled as $\sim$
4.5 arcmin FWHM Gaussians (\cite{kuo1}). A more extensive review of the
instrument and its capabilities can be found in Runyan et al. (2003). For
the 2001 season, observations were made at 150, 220, 275, and 345 GHz
(band widths of 30, 30, 50, and 25 GHz respectively), with a column of
four detectors per frequency.  These frequencies take full advantage of
the atmospheric windows available at the Pole. They bracket the null in
the thermal SZE spectrum and allow ACBAR to produce both SZE decrement and
SZE increment images. For the 2002 observing season, ACBAR was modified so
that each row performed observations at the same frequency. In addition,
the number of 150GHz filters was doubled, by removing the 345 GHz
channels.

		During an observation, the chopping motion of the tertiary mirror sweeps each bolometer beam across the sky in the azimuth direction at an approximately constant elevation. The design of the telescope keeps the illumination pattern on the primary almost constant during the chopping; however, this is not the case over the secondary mirror. This introduces a systematic offset into the data because the emissivity over the secondary is not constant due to snow accumulation, ice, and temperature variations. To remove these offsets, we used a Lead-Main-Trail technique. The method is as follows: The main field sweeps (where the target is centered) are led and trailed by similar sets of raster scans (lead and trail scans) of the sky at the same elevation but spaced in azimuth by half a degree. The lead and trail observations are then subtracted from the main observation to correct for the offset. Note that the observing time in the lead and trailed stares is half the exposure time of the lead stare (typically, the Lead and Trail fields are observed for 30 sec while the Main field is observed for 60 seconds). This guarantees that the L-M-T observing sequence is performed fast enough so that we minimize the effects of any temporal variation in the offsets. 

Note that for the cluster observations, we set the chopper throw to be
$\sim$ 1.5 degrees. With a sampling rate of $\sim$ 300 Hz, this produced.  
$\sim$ 200 samples per arcmin per stare. In order to make a 2d-map of a
cluster, we take several stares (set of L-M-T field observations) spaced
in elevation by 1 arcmin such that we cover the cluster in its entirety
(e.g., for a $z=$0.05 cluster, we map a region that is 1 degree in
elevation). In addition, we combine several (30 $\rightarrow$ 50) of these raster scans in order
to integrate long enough to build up signal to noise.

\section{Observation of Unresolved Galaxy Clusters}

During the 2001-2002 observing season we observed two $z \sim$ 0.3 clusters: Abell S1063 ($z=$0.347) and 1E0657-67 ($z=$0.299). These two clusters are the X-ray brightest clusters in the REFLEX catalog that could be observed from the South pole. We detected the thermal Sunyaev-Zeldovich 
effect from 1E0657-67 at 150 GHz and 275GHz (Figure 1). These are the first 2d-maps of the SZE at either side of the thermal null obtained for a cluster with the same instrument. Moreover, we also report a non-detection at 220 GHz consistent with the thermal SZE. This cluster was first discovered by the {\it Einstein} satellite (~\cite{1995ApJ...444..532T}). A ROSAT image of this cluster confirmed this previous detection and also revealed the presence of bimodality. Detailed {\it Chandra} X-ray and temperature maps derived by Markevitch et al. (\cite{2002ApJ...567L..27M}) showed that this cluster is a recent merger. In fact, the merger is so recent that the bulk shock produced by the core crossing is still strong (Mach number of $\sim$ 2-3) and there is significant temperature substructure in the system with temperature ranging from 8 keV to more than 20keV. Our observations revealed a peak decrement of $\sim$ 180 $u$K/beam (with an error of 9 $u$K/beam), a peak increment of $\sim$ 180 $u$K (with an error of 66 $u$K/beam), and a negligible signal at 220 GHz. Note that our 150GHz and 275GHz detections confirm an early SZE detection obtained by Andreani et al. (~\cite{1999ApJ...513...23A}). We have also detected, for the first time, a thermal SZE signature from Abell S1063. At 150 GHz the cluster was detected with a signal-to-noise of $\sim$ 3.5.

\begin{figure}[!p]
  \includegraphics[height=.99\textheight]{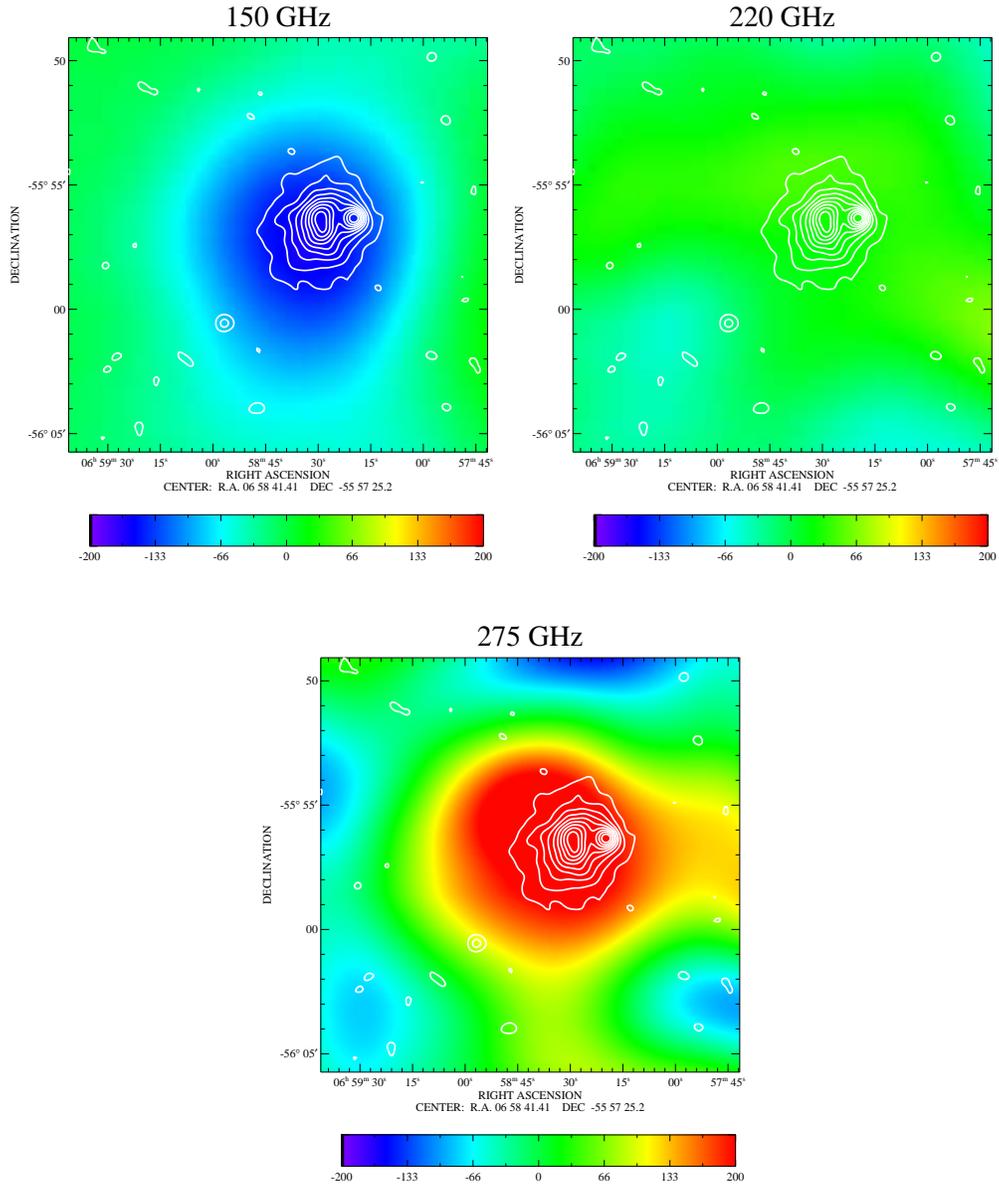}
  \caption{Colorscale 150GHz, 220GHz, and 275GHz ACBAR images of 1E0657-67 overlaid onto ROSAT HRI white contours. The ACBAR beam size is $\sim$ 4.5 arcmin. The horizontal colorbar shows the CMB temperature scale of the maps in $u$K and ranges from $-$200$u$K/beam to 200$u$K/beam. The noise in the central region of each image is $\sim$ 10$u$K/beam, 30$u$K/beam, and 70$u$K/beam respectively. Each map is $\sim$ 15 arcmin $\times$ 15 arcmin.}
\end{figure}

\section{Observation of Resolved Galaxy Clusters}

	In addition, we also observed the 5 brightest nearby ($z <$0.1) clusters accessible from the South Pole: Abell 3667, Abell 3266, Abell 3158, Abell 3827, and Abell 3112. They have been observed to different levels of sensitivity (typical noise in the central region at 150GHz are: 18$u$K/beam, 38$u$K/beam, 6$u$K/beam, 11$u$K/beam, and 19$u$K/beam respectively). 


As expected, the maps of nearby clusters ($z <$0.1) were contaminated by
primary CMB anisotropies. This contamination was especially pronounced in
the 150 GHz maps, where the CMB noise dominated the random noise.  However, our multifrequency observations and our
knowledge of the spectrum of the physical mechanism responsible for the thermal SZE allows us to spectrally filter out the CMB primary anisotropy (~\cite{gom1}).  Intuitively, this method is very simple as it can be shown that at 150 GHz we detect the CMBR minus the cluster signal; at 220 GHz -the thermal SZE null- the CMBR
dominates the map; and at 275 GHz we detect the CMBR plus the cluster
signal. So, naively subtracting the 150 GHz data from the 275 GHZ data would retrieve the cluster signal and cancel out the CMBR data. In practice, we construct a linear 
combination of the 150, 220, and 275 GHz images designed to maximize 
the SZ signal and minimize the CMBR signal at the RJ limit that takes into account the shape of the SZE spectrum. In 2004 we plan to complete our multi-frequency ACBAR observations of a sample of 10 nearby
clusters ($z <$ 0.1)


\section{Sunyaev-Zeldovich Effect Blind Cluster Survey}
To date, $\sim$ 10 deg$^2$ of sky have been mapped by ACBAR to a
sensitivity of 5uK/beam at 150 GHz. A further $\sim$ 10 deg$^2$ have been mapped to a
sensitivity of 11 uK/beam at 150 GHz.  These survey regions were centered
on the radio bright quasars, allowing us to register the individual raster
scans and refine the pointing model to within an arcminute. An analysis of the CMBR power spectrum using these data has been 
presented in  Kuo et al. 2003 (~\cite{kuo1}) and Goldstein et al. (~\cite{gol1}). We have also used 
these maps to search for previously undetected clusters of galaxies by applying an optimal filtering to remove CMBR and instrument noise (see Runyan 2003 ~\cite{run2}).

In this way, we have
uncovered more than 40 high significance 
cluster candidates in our survey fields (over 3$\sigma$). The
ACBAR detections differ from previous detections of the SZE at high
redshift as these have come from pointed searches toward known (~\cite{2003ApJ...583..559L}) clusters. The ACBAR
Blind Cluster Survey selection function is well understood, so we
should ultimately be able to use the ACBAR SZE clusters to constrain
cosmological parameters once candidates are confirmed and redshifts.

	We have performed an optical followup to these cluster candidates with the MOSAIC camera mounted on the 4m Blanco telescope at CTIO. We have obtained Sloan $g$, $r$, and $i$ images for most of the candidates and identified red-sequences in 6 of them that appear to be consistent with clusters at $z>$0.5. Once confirmed
spectroscopically, these will be the first
clusters to have been discovered via their SZ signal alone (rather than
via optical, radio or X-ray signatures).


\begin{theacknowledgments}
 We acknowledge the support of Center for Astrophysics Research in Antarctica (CARA) 
polar operations has been essential in the installation and operation of 
the telescope.  We would like to thank John Carlstrom and Steve Meyer
for their early and continued support of the project.
We gratefully acknowledge Simon Radford for 
providing the 350 $\mu$m tipper data.  We thank Charlie Kaminski
and Michael Whitehead for their assistance with winter observations.
We also would like to acknowledge the help and fruitful discussion with P. Ade, J. Bock, C. Cantalupo, M. Daub, E. Torbet, and C. Reichardt.
This work has been supported by NSF Office of Polar
Programs grants OPP-8920223 and OPP-0091840, by NASA through an LTSA grant NAG5-7926; the NASA XMM Guest Observer program; and through the American Astronomical Society's Small Research Grant Program, and by an AAS travel grant.
\end{theacknowledgments}


\bibliographystyle{aipproc}   

\bibliography{como4}

\IfFileExists{\jobname.bbl}{}
 {\typeout{}
  \typeout{******************************************}
  \typeout{** Please run "bibtex \jobname" to optain}
  \typeout{** the bibliography and then re-run LaTeX}
  \typeout{** twice to fix the references!}
  \typeout{******************************************}
  \typeout{}
 }

\end{document}